\documentclass[10pt,conference]{IEEEtran}

\usepackage{subcaption}
\usepackage{amssymb}
\usepackage{pifont}

\newcommand{\cmark}{\ding{51}}%
\newcommand{\xmark}{\ding{55}}%
\usepackage[flushleft]{threeparttable}
\usepackage[table,xcdraw]{xcolor}

\usepackage{subfig}
\usepackage{subfig}

\usepackage{amsmath,amssymb,amsfonts}
\usepackage{graphicx}
\usepackage{pifont}

\usepackage{multirow}

\usepackage[utf8]{inputenc} 
\usepackage{textcomp}
\usepackage{booktabs}
\usepackage[]{algorithm2e}
\usepackage{commath}
\usepackage{epstopdf}
\usepackage{float}
\usepackage{placeins}
\usepackage[round, sort, authoryear]{natbib}
\usepackage{mathrsfs}
\usepackage{multirow}
\usepackage{verbatim}
\usepackage{color}
\usepackage{url}
\usepackage{flushend}

\usepackage{wrapfig}
\providecommand{\AnhNTAuthor}{Tuan Anh Nguyen}

\providecommand{\AnhNTAuthorAffiliation}{%
  Konkuk Aerospace Design-Airworthiness Research Institute (KADA),
  Konkuk University, Seoul 05029, South Korea}

\providecommand{\DugkiMinAuthor}{Dugki Min}

\providecommand{\DugkiMinAuthorAffiliation}{%
  Department of Computer Science and Engineering, Konkuk University, Seoul 05029, South Korea}

\providecommand{\EunmiChoiAuthor}{Eunmi Choi}
\providecommand{\EunmiChoiAuthorEmail}{emchoi@kookmin.ac.kr}
\providecommand{\EunmiChoiAuthorAffiliation}{%
  School of Software, College of Computer Science,
  Kookmin University, Seoul 02707, South Korea}

  \providecommand{\AckProfessorLeeUAM}{This research was partially supported by Basic Science Research Program through the National Research Foundation of Korea (NRF) funded by the Ministry of Education(No. 2020R1A6A1A03046811).}
\providecommand{\AckProfessorMinMidsizedResearcher}{This research was supported by the Basic Science Research Program through the National Research Foundation of Korea (NRF) funded by the Ministry of Education (2021R1A2C2094943)}

\begin{document}
\IEEEoverridecommandlockouts

\title{Optimal Resource Utilization in Hyperledger Fabric: A Comprehensive SPN-Based Performance Evaluation Paradigm}

\author{%
Carlos Melo$^\mp$,
Glauber Gonçalves$^\mp$,  
Francisco A. Silva$^\mp$
Leonel Feitosa$^\mp$,
\\
Iure Fé $^\mp$,
André Soares$^\mp$, 
\EunmiChoiAuthor$^\divideontimes$,
\AnhNTAuthor$^\circledast$,
and 
\DugkiMinAuthor$^\dagger$, 
  \\
  $^\mp$Applied Research on Distributed Systems Lab (PASID), Federal University of Piauí (UFPI), Brazil \\
  $^\circledast$ \AnhNTAuthorAffiliation 
  \\
  $^\divideontimes$ \EunmiChoiAuthorAffiliation
  \\
  $^\dagger$ \DugkiMinAuthorAffiliation
  \\  
  Corresponding authors: \{anhnt2407, dkmin\}@konkuk.ac.kr
  \thanks{E-mails: \{leonelfeitosa, andre.soares, ggoncalves, faps, casm\}@ufpi.edu.br, \{anhnt2407, dkmin\}@konkuk.ac.kr, \EunmiChoiAuthorEmail} 
  \thanks{\AckProfessorLeeUAM; \AckProfessorMinMidsizedResearcher}
}

\IEEEpubid{\makebox[\columnwidth]{
978-1-7281-8086-1/20/\$31.00 \copyright2023 IEEE\hfill
} \hspace{\columnsep}\makebox[\columnwidth]{ }}

\maketitle

\begin{abstract}
Hyperledger Fabric stands as a leading framework for permissioned block-chain systems, ensuring data security and audit-ability for enterprise applications. As applications on this platform grow, understanding its complex configuration concerning various block-chain parameters becomes vital. These configurations significantly affect the system's performance and cost. In this research, we introduce a Stochastic Petri Net (SPN) model to analyze Hyper-ledger Fabric's performance, considering variations in block-chain parameters, computational resources, and transaction rates. We provide case studies to validate the utility of our model, aiding block-chain administrators in determining optimal configurations for their applications. A key observation from our model highlights the block size's role in system response time. We noted an increased mean response time, between 1 to 25 seconds, due to variations in transaction arrival rates.
\end{abstract}

\begin{IEEEkeywords}
blockchain, Petri net, performance evaluation, capacity planning.
\end{IEEEkeywords}

\section{Introduction}
\label{sec:introduction}

Blockchain, heralded as a transformative technology, promises significant potential in various sectors through secure, decentralized data storage \citep{xu_springer_book2019}, achieving transaction immutability, audibility, and consistency via asymmetric cryptography, consensus protocols, and peer-to-peer networks \citep{greve_minicurso_sbrc2018}. To facilitate broader industrial adoption, there's a pivot towards private or permissioned blockchain architectures like Hyperledger Fabric, which enhance performance and security for consortium members across sectors, including healthcare \citep{aguiar_fgcs2022} and finance \citep{spunta_2020}. These networks focus on reducing data recording latency and increasing transaction throughput, with Hyperledger Fabric's performance contingent on infrastructure and blockchain configurations, notably block size and \textit{timeout} for block creation.

This study introduces a Stochastic Petri Net (SPN) model to examine Hyperledger Fabric's performance across various configurations, addressing gaps in existing models by considering endorsement, ordering, and the \textit{commit} phase. Our model, validated through case studies, aims to refine Hyperledger Fabric configurations for enhanced industry adoption, analyzing how blockchain configurations and computational capabilities impact transaction times and throughput.

Contributions of this work include:
\begin{itemize}
    \item Developing an analytical model for Hyperledger Fabric blockchain configuration, providing operational efficiency insights before deployment;
    \item Conducting a detailed performance evaluation to identify transaction phase bottlenecks;
    \item Offering case studies as practical application evidence, complemented by a sensitivity analysis to identify key factors affecting Mean Response Time (MRT).
\end{itemize}
The paper progresses from reviewing relevant literature (Section \ref{sec:rel}), describing the SPN performance model and its importance in Hyperledger Fabric transactions (Section \ref{sec:model}), to showcasing case studies for insight (Section \ref{sec:results}), and concluding with future research avenues (Section \ref{sec:conclusion}).

\section{Literature Review}
\label{sec:rel}

Recent advancements in blockchain research have utilized various modeling techniques to analyze the performance and behavior of blockchain deployment architectures. Melo et al.~\citep{melo_computing2022,melo_supercomp2021} employed Continuous Time Markov Chains (CTMC), Reliability Block Diagrams (RBD), and Stochastic Petri Nets (SPN) to assess the availability and resource provisioning for Hyperledger Fabric and Ethereum platforms, aiding in strategic blockchain application planning. Studies by Jiang et al.~\citep{jiang_springer_p2pna2020}, Wu et al.~\citep{wu_acm_ease2022}, and Ke and Park~\citep{ke_springer_cc2022} have concentrated on Hyperledger Fabric's performance metrics, with models focusing on throughput, discard rates, queue lengths, and mean response times. Additional research by Xu et al.~\citep{xu_ipm2021}, Yuan et al.~\citep{yuan2020performance}, and Sukhwani et al.~\citep{sukhwani2018performance} explored throughput and latency effects through analytical models and Generalized Stochastic Petri Nets (GSPN), and Stochastic Reward Networks (SRN), respectively, to analyze transaction steps.

Our contribution is an SPN model that evaluates Hyperledger Fabric's performance metrics, including \textit{maximum block size rate} and \textit{timeout block rate}, alongside a calibration approach for these parameters. Our work, summarized in Table~\ref{tab:related}, provides a detailed examination of timeout and block size effects, especially dissecting the ordering phase to highlight block formation issues related to timeouts, offering a more nuanced understanding compared to previous studies.

\begin{table}[!htp]
\centering
\footnotesize
\setlength{\tabcolsep}{3pt}  
\begin{center}
\caption{Related Works}
\begin{tabular}{@{}lcccccc@{}}
\toprule
\rowcolor[HTML]{C0C0C0} 
\multicolumn{1}{c}{\textbf{Publication}} & \multicolumn{1}{c}{\textbf{MRT}} & \multicolumn{1}{c}{\textbf{Throughput}} & \multicolumn{1}{c}{\textbf{RU}} & \multicolumn{1}{c}{\textbf{MSBR}} & \multicolumn{1}{c}{\textbf{TBR}} & \multicolumn{1}{c}{\textbf{DTP}} \\ \midrule
\citep{sukhwani2018performance} & \cmark & \cmark & \cmark & \cmark & \cmark & \xmark \\
\citep{jiang_springer_p2pna2020} & \cmark & \cmark & \xmark & \xmark & \xmark & \cmark \\
\citep{yuan2020performance} & \cmark & \cmark & \xmark & \cmark & \cmark & \xmark \\
\citep{xu_ipm2021} & \cmark & \xmark & \xmark& \cmark  & \xmark & \xmark \\
\citep{melo_supercomp2021} & \xmark & \xmark& \cmark& \xmark  & \xmark & \xmark \\
\citep{ke_springer_cc2022} & \cmark & \xmark& \xmark& \xmark  & \xmark & \xmark \\
\citep{wu_acm_ease2022} & \cmark & \xmark& \xmark& \xmark  & \xmark & \xmark \\
\citep{melo_computing2022} & \xmark & \xmark& \cmark& \xmark  & \xmark & \xmark \\ \bottomrule
\end{tabular}
\label{tab:related}
\end{center}
\begin{tablenotes}[flushleft]\footnotesize
\item[]Metrics abbreviations: Mean Response Time (MRT), Resource Utilization (RU), Maximum Size Block Rate (MSBR), Timeout Block Rate (TBR), Discarding transactions probability (DTP)
 \par
\end{tablenotes}
\end{table}

\section{System Architecture}
~\label{sec:back}

\subsection{Hyperledger Fabric Framework}\label{subsec:arq}
\begin{figure}[!htb]
 \center
 \includegraphics[width=0.5\textwidth]{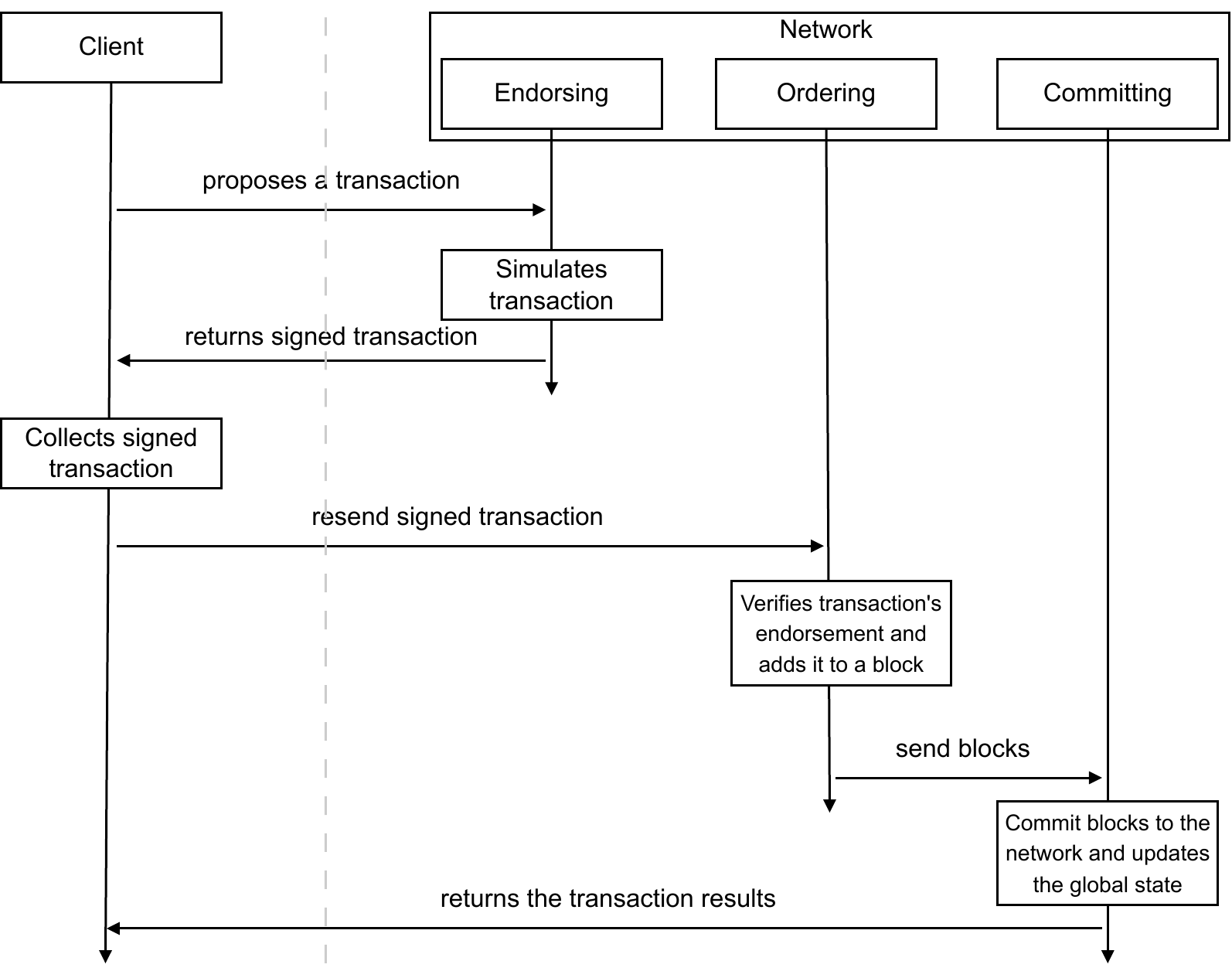}
 \caption{Transaction's flow in the Hyperledger Fabric platform.}
 \label{fig:arch:hlf} 
\end{figure}

The Hyperledger Fabric (HLF) has emerged as a leading blockchain platform, supported by over 35 institutions and more than 200 developers.\footnote{For more details, see \textit{https://hyperledger-fabric.readthedocs.io}.} Unlike public blockchains such as Bitcoin and Ethereum, HLF adopts an execute-order-validate model for transactions to improve scalability~\citep{androulaki_eurosys2018}. It allows for diverse configurations, primarily between clients and service providers, with the latter significantly impacting system performance. HLF leverages container technologies for managing "chaincodes," or smart contracts, which encapsulate the business logic necessary for deployment, while clients interact through SDKs in languages like Javascript or Python. The HLF transaction pathway involves endorsement, ordering, and commit phases, functioning across multiple network nodes as depicted in Figure~\ref{fig:arch:hlf}, showcasing the integral steps from client initiation to network processing.

\begin{figure}[!htb]
 \center
 \includegraphics[width=0.35\textwidth]{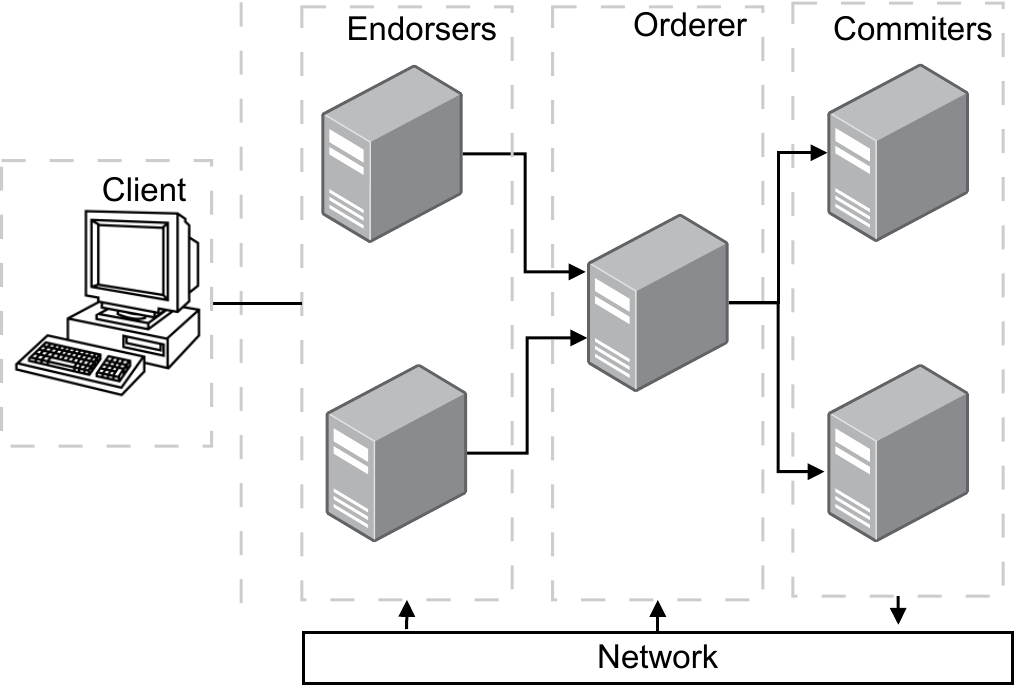}
 \caption{Architecture Overview.}
 \label{fig:architecture} 
\end{figure}

The transaction process in Hyperledger Fabric initiates with a client application submitting a proposal to a network node, followed by seeking endorsements from more than half of the nodes. Post-endorsement, the transaction progresses to the \textit{ordering} phase, where it awaits the formation of a new block upon hitting a preset timeout or block size. During the \textit{endorsement} phase, the transaction undergoes simulation, receiving approval or rejection. Subsequently, blocks enter the \textit{commit} phase for validation and integration into the blockchain, updating the global state for swift queries. Illustrated in Figure \ref{fig:architecture} is a typical configuration featuring two endorsing nodes, one ordering node, and two committing nodes, showcasing the system's scalability and the significance of modeling for optimizing pre-deployment performance and cost efficiency.

\section{Base Model}
~\label{sec:model}

\begin{figure}[!htb]
    \centering
        \includegraphics[width=0.35\textwidth]{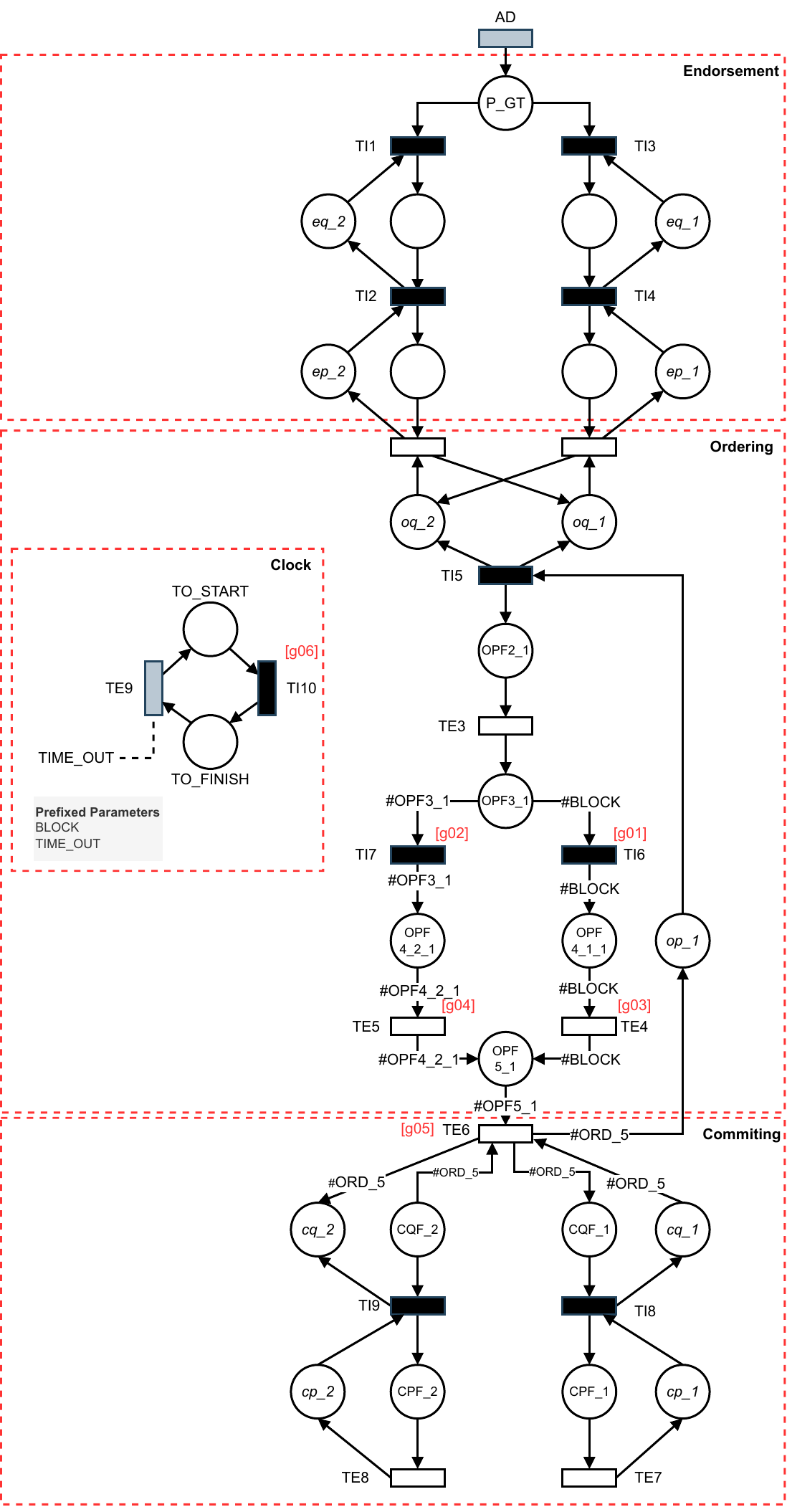}
     \caption{SPN model for processing transactions on Hyperledger Fabric.}
  \label{fig:model}
\end{figure}

Our research presents a Stochastic Petri Nets (SPN) model designed to evaluate Hyperledger Fabric's performance, a permissioned blockchain. This model enables administrators to adjust parameters optimally before real-world deployment. Illustrated in Figure~\ref{fig:model}, the SPN model simulates transactions through Hyperledger Fabric's three principal phases: \textit{endorsement, ordering}, and \textit{commit}, spanning across $N$ nodes. For simplification, our model posits that each phase occurs on two nodes, with the exception of the ordering phase, which is managed by a single node. Transaction initiation by an application and its consistent arrival rate are depicted by a deterministic gray transition, labeled as the arrival delay (\texttt{AD}).

In the Hyperledger Fabric framework, the transaction lifecycle commences with the \textit{endorsement} phase, where a token at \texttt{P\_GT} initiates processing on one of two computational nodes via transitions \texttt{T11} or \texttt{T13}. Nodes, represented by two triangles for input and processing queues, operate differently: the input queue transitions instantly, while the endorsement process is subject to timed transitions \texttt{TE1} and \texttt{TE2}, reflecting node capacities like \texttt{EP\_1} and leading to queue formations when exceeded. The \textit{ordering} phase, handled by a single node, follows with a similar queue setup. It sorts transactions based on \texttt{op\_1} capacity, with initial processing marked by \texttt{TE3} for block formation. Tokens at \texttt{OPF3\_1} lead to either complete or partial blocks, depending on the \texttt{\#BLOCK} parameter or a timeout. These blocks then proceed through \texttt{TE4} and \texttt{TE5}. Upon reaching \texttt{OPF5\_1}, the system's \texttt{Clock} is reset. The \textit{commit} phase is triggered by transition \texttt{TE6}, involving all designated nodes and culminating with \texttt{TE7} and \texttt{TE8}. The model's components and operational conditions are elaborated in Tables \ref{tab:descrimodelo} and \ref{tab:condigruarda}, respectively.

\begin{table}[]
\centering
\begin{tabular}{@{}lll@{}}
\toprule
\rowcolor[HTML]{C0C0C0} 
\multicolumn{1}{c}{\cellcolor[HTML]{C0C0C0}\textbf{Type}} & \multicolumn{1}{c}{\cellcolor[HTML]{C0C0C0}\textbf{Element}} & \multicolumn{1}{c}{\cellcolor[HTML]{C0C0C0}\textbf{Description}} \\ \midrule
\textbf{Places} & P\_GT    & Wait for new transactions  \\ \midrule
\textbf{Transitions}      & AD & Transactins Arrival Time  \\
\textbf{Deterministic}    & TE9& TIME\_OUT assigned time\\ \midrule & TE1, TE2 & Endorsement time       \\
& TE3& Block assembly time    \\
& TE4, TE5 & Block processing time  \\
& TE6& Commit process         \\
\multirow{-5}{*}{\textbf{Timed Transitions}}    & TE7, TE8 & Time for committing        \\ \midrule
& TI1, TI3 & Endorsement queue      \\
& TI2, TI4 & Endorsement process    \\
& TI5& Ordering queue         \\
& TI6& Ordering due to block size  \\
& TI7& Ordering due to TIME\_OUT       \\
& TI8, TI9 & Commit Process         \\
\multirow{-7}{*}{\textbf{Immediate Transitions}}& TI10     & TIME\_OUT's restart    \\ \midrule
& eq1\_1, eq1\_2     & Endorsement queue capacity       \\
& ep1\_1, ep1\_2     & Endorsement process capacity     \\
& oq\_1    & Ordering queue's capacity        \\
& op\_1    & Ordering process capacity        \\
& cq1\_1, cq1\_2     & Commit queue's capacity\\
\multirow{-6}{*}{\textbf{Token's Places}} & cp1\_1, cp1\_2     & Commit process capacity\\ \bottomrule
\end{tabular}
\caption{Model's elements description, including places and transitions.}
\label{tab:descrimodelo}
\end{table}

\begin{table}[]
\centering
\begin{tabular}{@{}lll@{}}
\toprule
\rowcolor[HTML]{C0C0C0} 
\multicolumn{1}{c}{\cellcolor[HTML]{C0C0C0}\textbf{Trans.}} &
  \multicolumn{1}{c}{\cellcolor[HTML]{C0C0C0}\textbf{ID}} &
  \multicolumn{1}{c}{\cellcolor[HTML]{C0C0C0}\textbf{Guard Expression}} \\ \midrule
TI6  & g01 & \#OPF3\_1\textgreater 0                    \\
TI7  & g02 & \#TO\_FINISH=1)AND(\#OPF3\_1\textgreater 0) \\
TE4  & g03 & \#OPF4\_1\_1\textgreater 0                 \\
TE5  & g04 & \#OPF4\_1\_2\textgreater 0                 \\
TE6  & g05 & \#OPF5\_1\textgreater 0                    \\
TI10 & g06 & \#OPF5\_1\textgreater 0                    \\ \bottomrule
\end{tabular}
\caption{Guard expressions for transitions (Trans.) whose indices (ID) are highlighted in red in the model.}
\label{tab:condigruarda}
\end{table}


\subsection{Metrics}

The mean response time (\textit{MRT}) is derived using Little's Law\citep{jainart} as $\frac{\texttt{TransactionsInProgress}}{\texttt{Arrival\_Rate}}$, correlating the system's average transactions in progress with the arrival rate, where $Arrival\_Rate = \frac{1}{Arrival\_Delay}$. This calculation presupposes a stable system, where the transaction rate does not exceed the server's processing capability. To determine the system's transactions in progress, we sum the tokens in places representing ongoing transactions, using $E(Place)$ for the expected number of tokens per place, calculated as $E(Place) = (\sum_{i=1}^{n} P( m(Place)=i)\times i)$. The discard rate probability is determined by the likelihood of $n$ tokens in a place being zero, expressed as $\texttt{P((EQ\_1=0)}\wedge \texttt{(EQ\_2=0))}$, with $\wedge$ signifying a logical AND. Resource utilization is assessed by comparing the expected tokens in a queue or process to its capacity, exemplified by the endorsement step's processing utilization as $\frac{\texttt{E(EP\_1)}}{\texttt{ep\_1}}$. System throughput, particularly at the \textit{commit} phase performed across $N$ computers, is the average exit rate, calculated by dividing the expected tokens by the service time, $\frac{\texttt{E(CPF\_1)}}{\texttt{t(TE7)}}$. Forking paths from \texttt{OPF3\_1} determine transition rates for timeouts or complete blocks, adhering to the aforementioned calculation principles. Additional metrics, BLOCK\_CALL\_RATE and TIME\_OUT\_CALL\_RATE, are detailed in equations \ref{eq:c1} and \ref{eq:c2}, respectively.

\begin{equation}\label{eq:c1}
\footnotesize
	\begin{aligned}
		\textbf{BLOCK\_CALL\_RATE} = \frac{\texttt{E(OPF\_1\_1)}}{\texttt{t(TE4)}}
	\end{aligned}
\end{equation}

\begin{equation}\label{eq:c2}
\footnotesize
	\begin{aligned}
		\textbf{TIME\_OUT\_CALL\_RATE} = \frac{\texttt{E(OPF\_2\_1)}}{\texttt{t(TE5)}}
	\end{aligned}
\end{equation}

\section{Results}~\label{sec:results}
\label{p3}
This section presents four case studies, each focusing on distinct parameters to demonstrate our model's effectiveness and its applicability to real-world systems. Using the Mercury tool~\citep{maciel2017mercury} for model representation and numerical analysis, we maintain the architecture and computational node setup as described in Section~\ref{sec:model}. Initial model parameters, detailed in Table~\ref{tab:cenarioC}, include queue capacity (cq), processing capacities for the endorsement (ep), ordering (op), and commit (cp) phases, along with service times for transitions (TE). These parameters, informed by Hyperledger platform testing, reflect computational nodes' processing capabilities and total memory allocations for queue capacities in each phase, highlighting that parameter values may differ based on the computational environment examined.

\begin{table}[]
\centering
\caption{Initial model configuration parameters used in case studies.}
\begin{tabular}{@{}llr@{}}
\toprule
\rowcolor[HTML]{C0C0C0} 
\multicolumn{1}{c}{\cellcolor[HTML]{C0C0C0}\textbf{Type}} &
  \multicolumn{1}{c}{\cellcolor[HTML]{C0C0C0}\textbf{Parameters}} &
  \multicolumn{1}{c}{\cellcolor[HTML]{C0C0C0}\textbf{Values}} \\ \midrule
                           & ep\_1, ep\_2, op\_1, cp\_1, cp\_2 & 6     \\
\multirow{-2}{*}{Capacity} & eq\_1, eq\_2, oq\_1, cq\_1, cq\_2 & 100   \\ \midrule
                           & TE1, TE2, TE3                     & 5 ms  \\
                           & TE6                               & 10 ms \\
                           & TE7, TE8                          & 80 ms \\
\multirow{-4}{*}{Time}     & TE4, TE5                          & 2 ms  \\ \bottomrule
\end{tabular}
\label{tab:cenarioC}
\end{table}


\subsection{Case Study 01 - Committer Capacity}

In our first case study, we examined the impact of varying processing capacities (\texttt{cp\_1} and \texttt{cp\_2}) during the \textit{commit} phase, testing configurations with each node's capacity set to 2, 4, and 6 containers. To comprehensively understand the effects, we also adjusted the transaction arrival rate from a low of 0.0025 requests per millisecond (req/ms) to a high of 0.3 req/ms, in increments of 0.01565 req/ms. For analytical clarity, we opted for a configuration where each block contained a single transaction, setting a long \textit{timeout} (\texttt{TIME\_OUT} = 10000ms) and a minimal block size (\texttt{BLOCK} = 1). This approach primarily focused the model's dynamics on the generation of complete blocks.

\begin{figure*}[!htb]
\centering
\begin{subfigure}[b]{0.25\linewidth}
  \includegraphics[width=\linewidth]{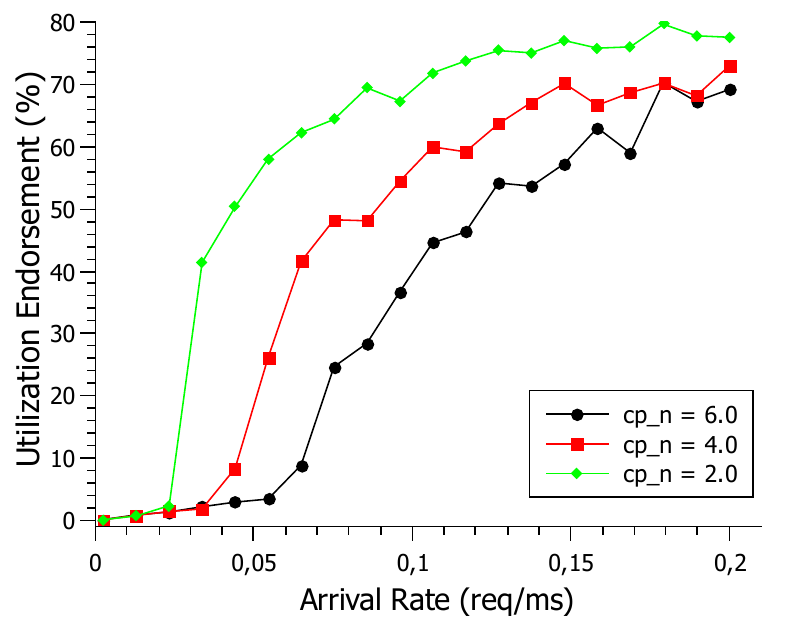}
  \caption{Usage - Endorsement}
  \label{fig:AR_U_end}
\end{subfigure}%
\hfill
\begin{subfigure}[b]{0.25\linewidth}
  \includegraphics[width=\linewidth]{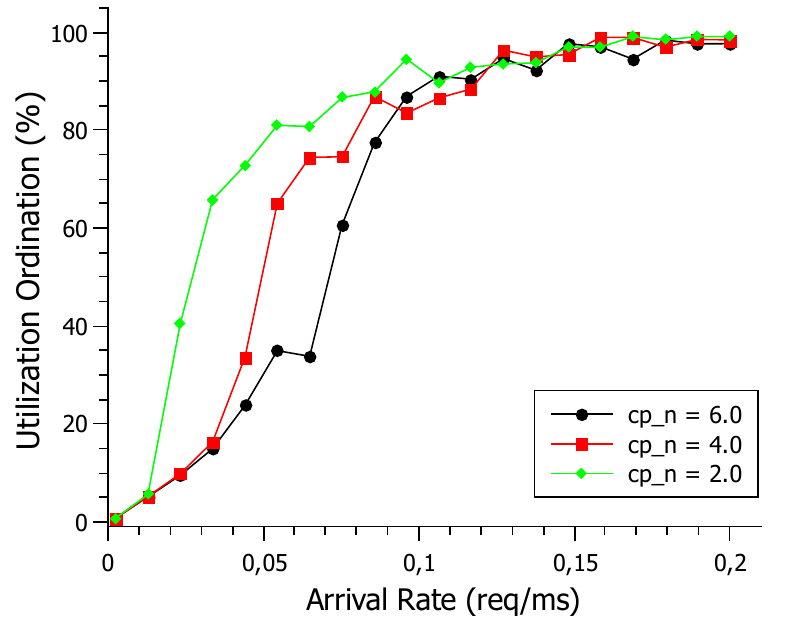}
  \caption{Usage - Ordering}
  \label{fig:AR_U_Ord}
\end{subfigure}%
\hfill
\begin{subfigure}[b]{0.25\linewidth}
  \includegraphics[width=\linewidth]{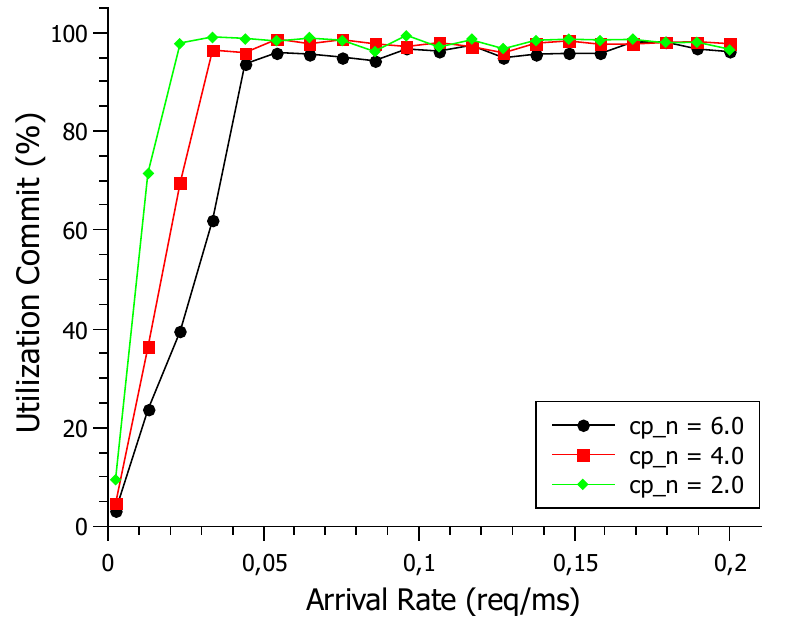}
  \caption{Usage - Commit}
  \label{fig:AR_U_Com}
\end{subfigure}%
\hfill
\begin{subfigure}[b]{0.25\linewidth}
  \includegraphics[width=\linewidth]{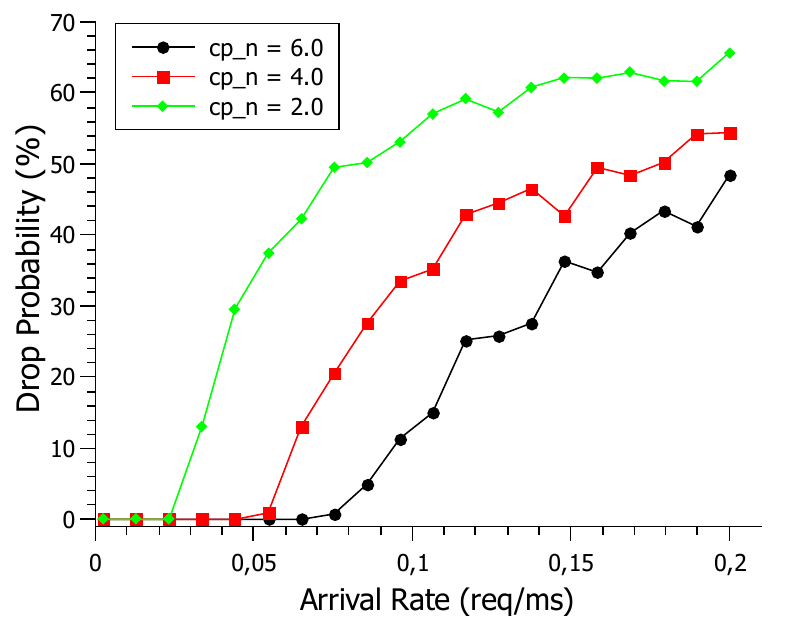}
  \caption{Discard Rate}
  \label{fig:AR_DP}
\end{subfigure}%
\hfill
\begin{subfigure}[b]{0.25\linewidth}
  \includegraphics[width=\linewidth]{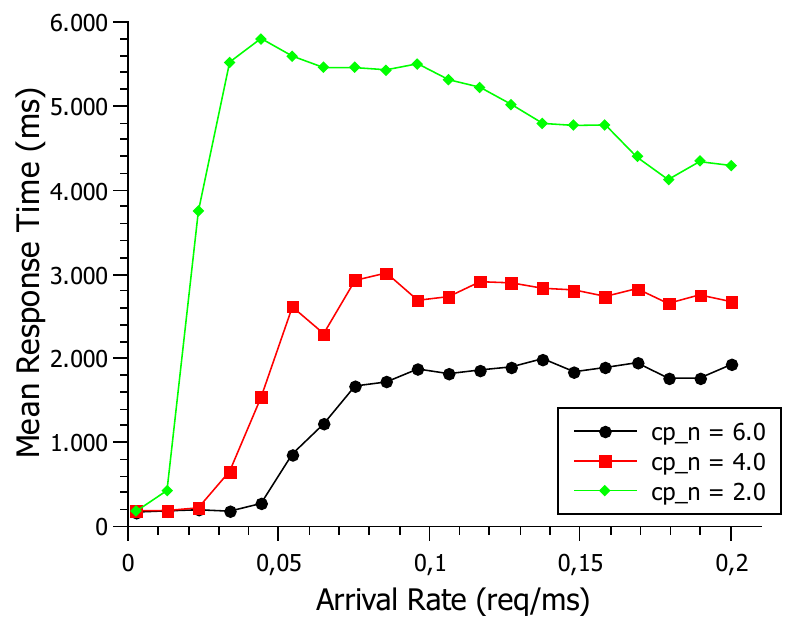}
  \caption{MRT}
  \label{fig:AR_MRT}
\end{subfigure}%
\hfill
\begin{subfigure}[b]{0.25\linewidth}
  \includegraphics[width=\linewidth]{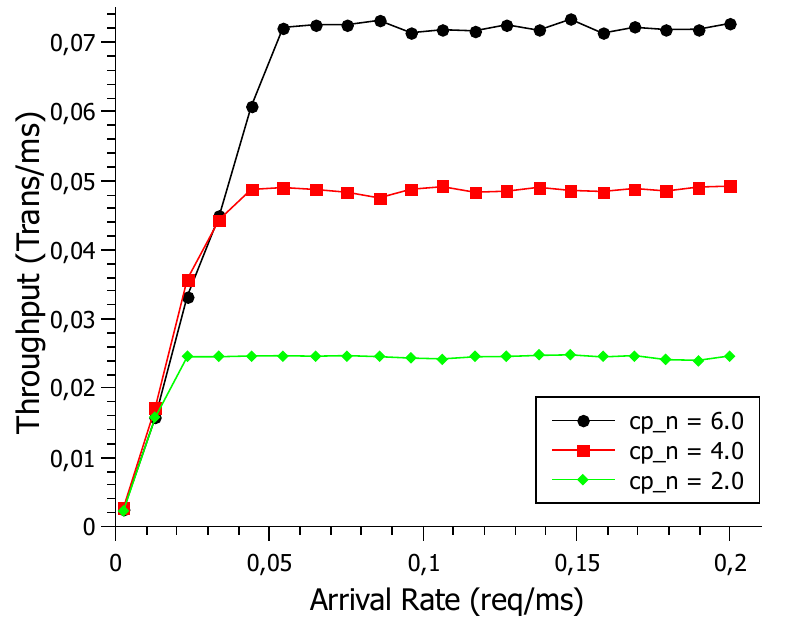}
  \caption{Flow}
  \label{fig:AR_TP}
\end{subfigure}
\hfill
\begin{subfigure}[b]{0.25\linewidth}
  \includegraphics[width=\linewidth]{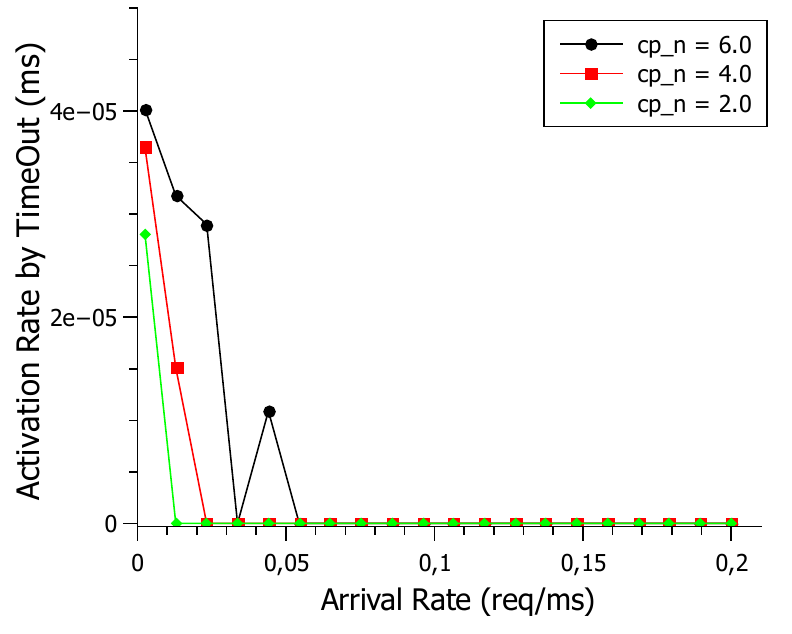}
  \caption{Timeout Activation}
  \label{fig:AR_TCALL}
\end{subfigure}%
\hfill
\begin{subfigure}[b]{0.24\linewidth}
  \includegraphics[width=\linewidth]{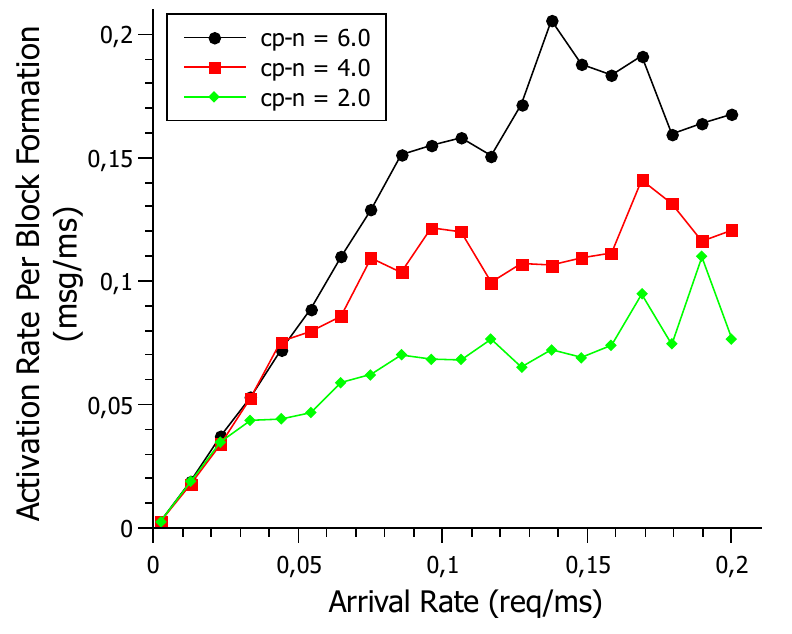}
  \caption{Size Activation Block}
  \label{fig:AR_BCALL}
\end{subfigure}
\caption{Case Study 01 - Committer Capacity}
\label{fig:sc1}
\end{figure*}

Figure \ref{fig:sc1} presents the outcomes of our first case study through eight distinct plots, highlighting the effects of parameter changes on system performance.

Utilization trends across Figures \ref{fig:sc1}a, \ref{fig:sc1}b, and \ref{fig:sc1}c show uniform increases with rising arrival rates, with \textit{commit} phase utilization notably accelerating as computational capacities are adjusted. An important observation is that beyond a 0.05 req/ms arrival rate in the \textit{commit} phase, changing capacities no longer impacts utilization, indicating a threshold for optimizing hardware investments.

The system's discard rate at the endorsement phase (Figure \ref{fig:sc1}d) parallels the endorsement utilization graph, suggesting that high utilization at this stage leads to increased discard probabilities due to congestion.

Mean Response Time (MRT) analysis (Figure \ref{fig:sc1}e) reveals performance degradation with increasing arrival rates, especially for a capacity (\texttt{cp\_n}) of 2. For capacities of 4 and 6, a significant MRT difference emerges at a 0.05 req/ms arrival rate, indicating the effect of capacity on response times.

The system's flow behavior (Figure \ref{fig:sc1}f) initially increases then stabilizes, limited by \textit{commit} phase utilization saturation, indicating a cap on performance gains beyond certain utilization levels.

Timeout activation rates (Figure \ref{fig:sc1}g) mostly remain negligible due to the extended 10000ms timeout, favoring complete block formation. Conversely, block activation rates (Figure \ref{fig:sc1}h) are expected to correlate with the arrival rate, but are constrained at lower \texttt{cp\_n} values, reflecting limited throughput due to restrictive capacities.

This case study underscores the \textit{commit} phase's critical role in system performance, demonstrating that optimal capacity tuning is essential to prevent performance drops. Despite increased capacities, the similarity in MRT for \texttt{cp\_n} values of 4 and 6 suggests that a \texttt{cp\_n} of 4 could be adequate for the studied arrival rates, offering an opportunity for resource optimization.

\subsection{Case Study 02 - Evaluating the Effects of Block Size and Timeout}

\begin{figure*}[!htb]
\centering
\begin{subfigure}[b]{0.24\linewidth}
  \includegraphics[width=\linewidth]{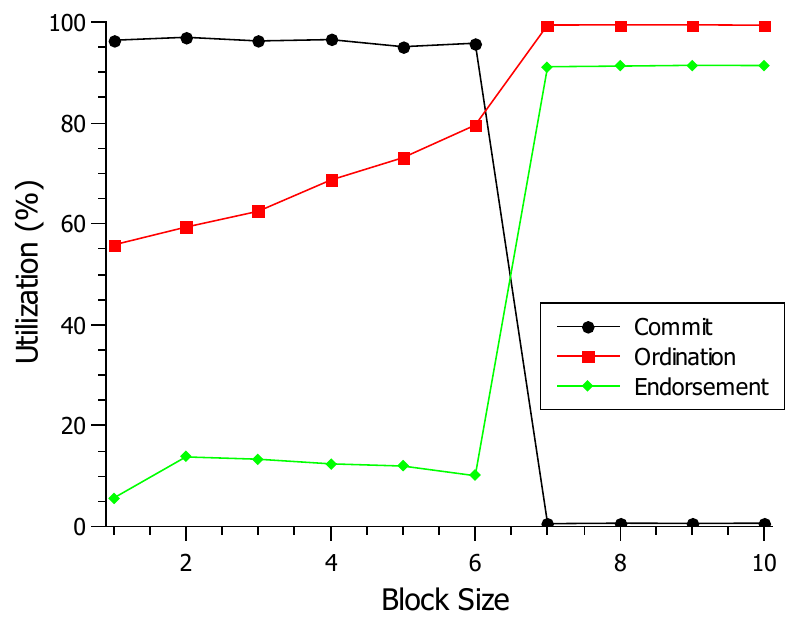}
  \caption{Usage by Changing Block}
  \label{fig:Blo_U}
\end{subfigure}%
\hfill
\begin{subfigure}[b]{0.24\linewidth}
  \includegraphics[width=\linewidth]{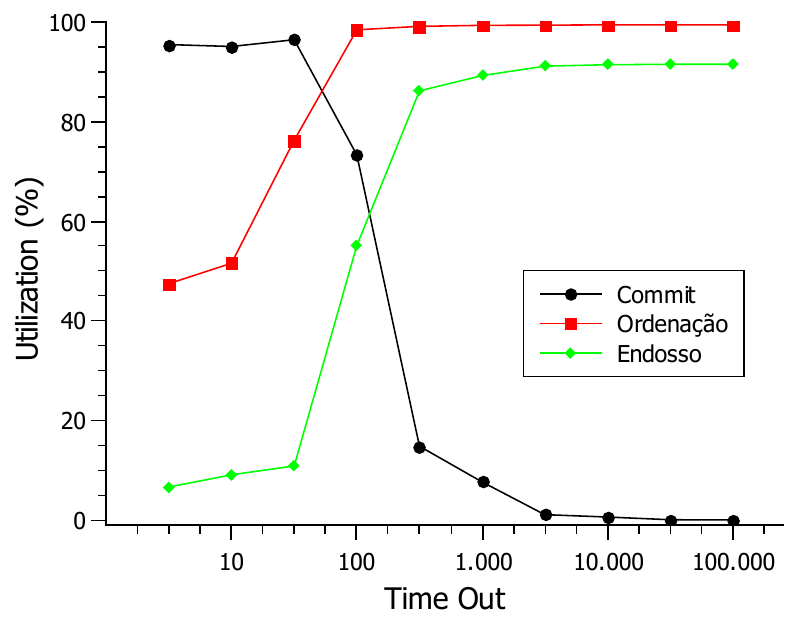}
  \caption{Usage by Changing Timeout}
  \label{fig:Tim_U}
\end{subfigure}%
\hfill
\begin{subfigure}[b]{0.24\linewidth}
  \includegraphics[width=\linewidth]{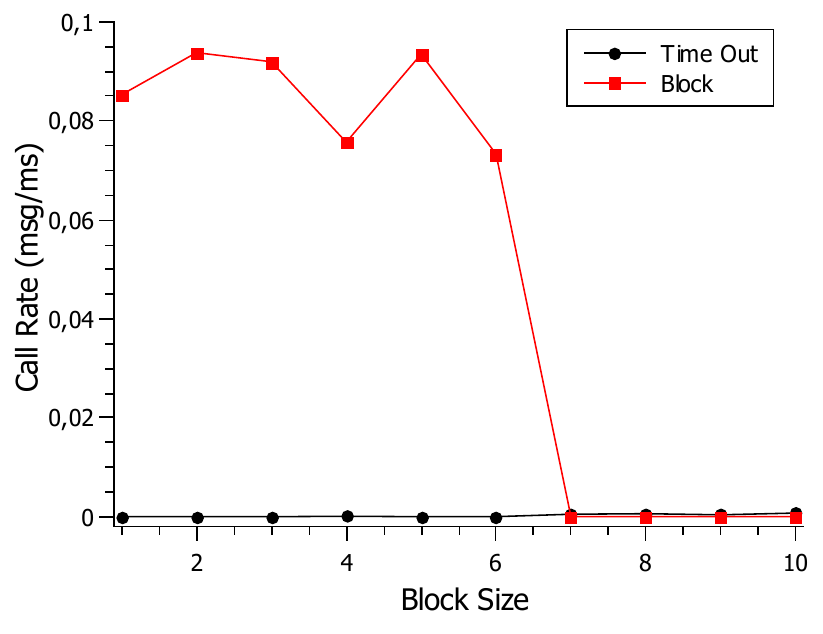}
  \caption{Block size Activation}
  \label{fig:Blo_Call}
\end{subfigure}%
\hfill
\begin{subfigure}[b]{0.24\linewidth}
  \includegraphics[width=\linewidth]{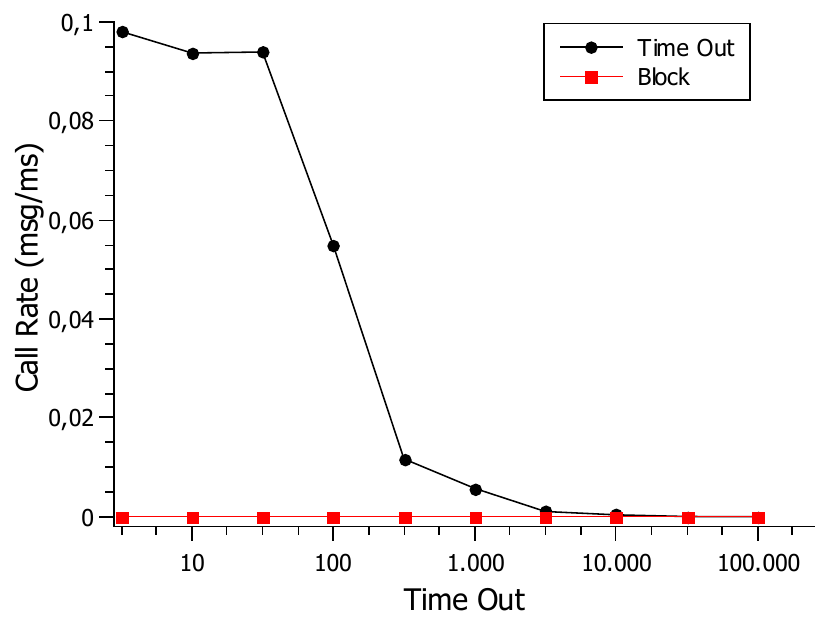}
  \caption{Timeout Activation}
  \label{figTim_Call}
\end{subfigure}%
\hfill
\begin{subfigure}[b]{0.24\linewidth}
  \includegraphics[width=\linewidth]{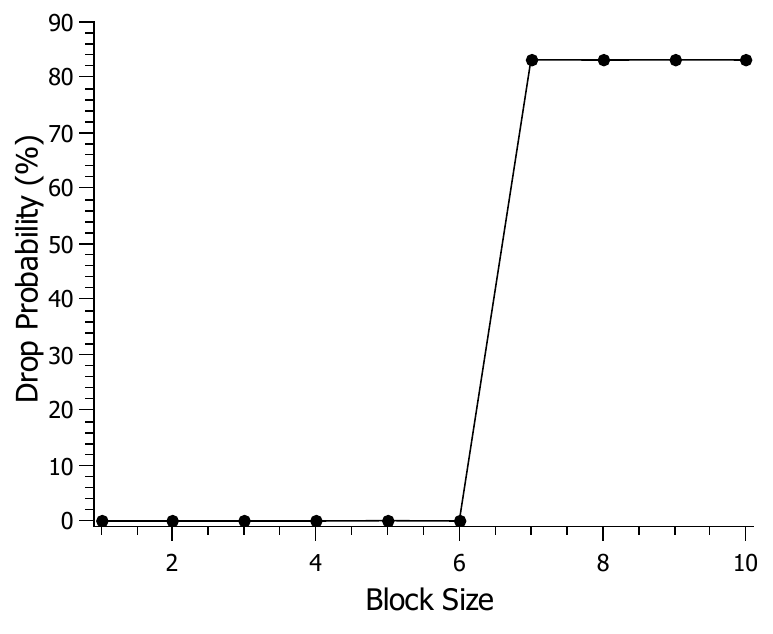}
  \caption{Drop Rate - Changing the Block}
  \label{fig:Blo_DP}
\end{subfigure}%
\hfill
\begin{subfigure}[b]{0.24\linewidth}
  \includegraphics[width=\linewidth]{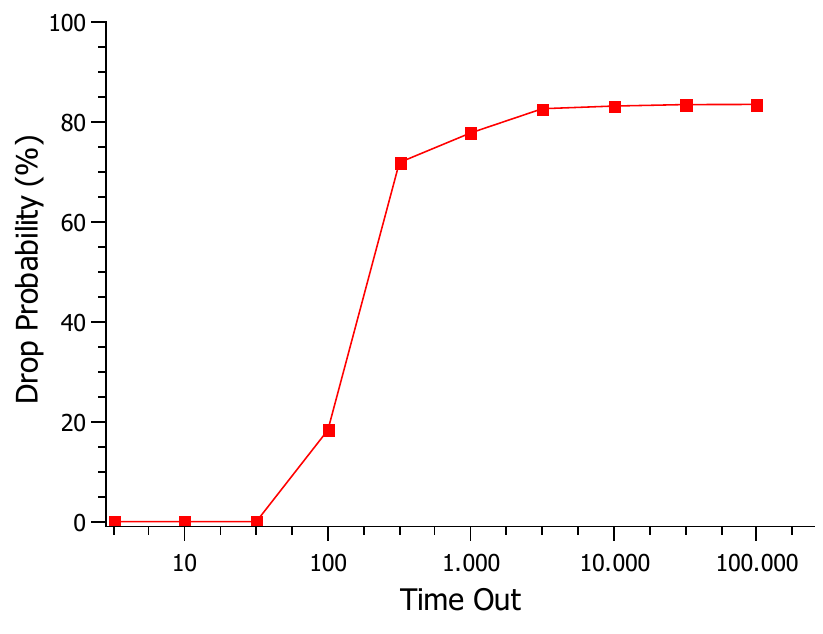}
  \caption{Drop Rate - Changing Timeout}
  \label{fig:Tim_DP}
\end{subfigure}%
\hfill
\begin{subfigure}[b]{0.24\linewidth}
  \includegraphics[width=\linewidth]{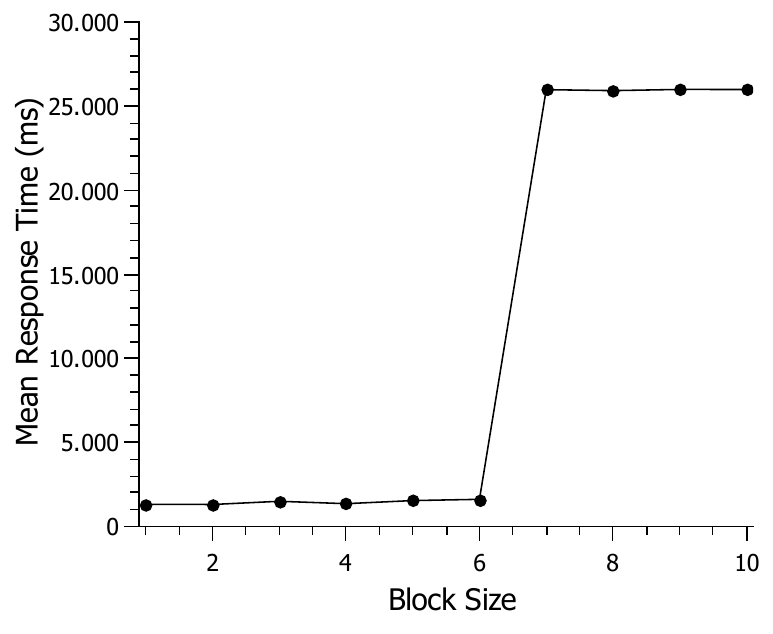}
  \caption{MRT - Changing Block}
  \label{fig:Blo_MRT}
\end{subfigure}%
\hfill
\begin{subfigure}[b]{0.24\linewidth}
  \includegraphics[width=\linewidth]{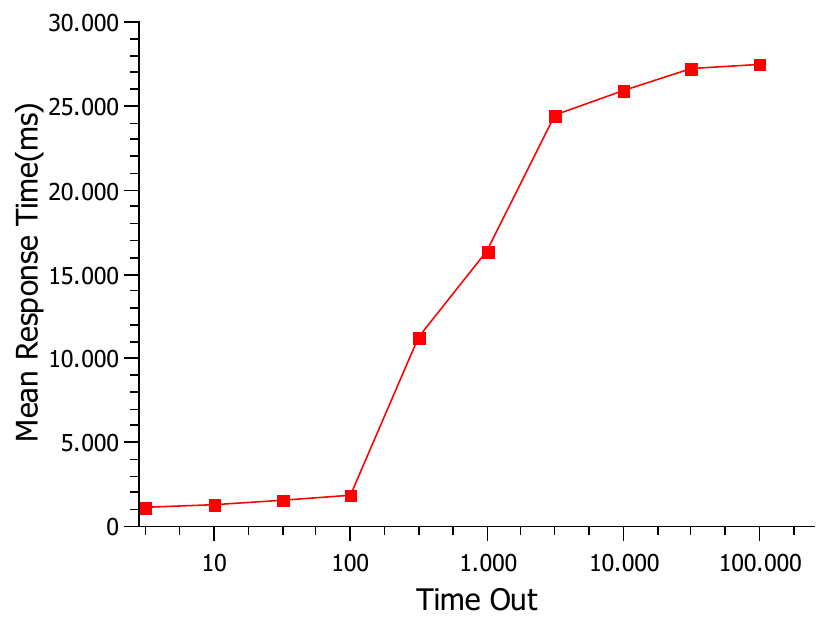}
  \caption{MRT - Changing Timeout}
  \label{fig:Tim_MRT}
\end{subfigure}%
\hfill
\begin{subfigure}[b]{0.24\linewidth}
  \includegraphics[width=\linewidth]{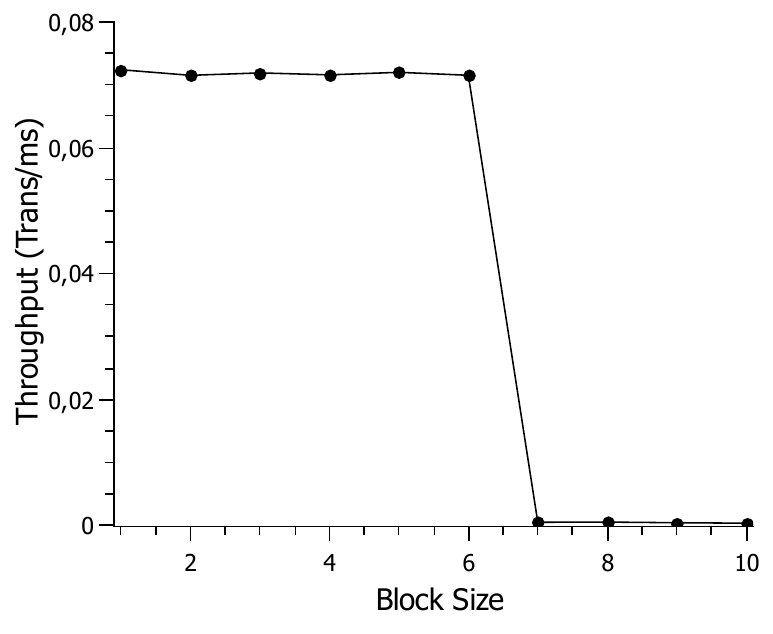}
  \caption{Flow - Changing Block}
  \label{fig:Blo_TP}
\end{subfigure}%
\hfill
\begin{subfigure}[b]{0.24\linewidth}
  \includegraphics[width=\linewidth]{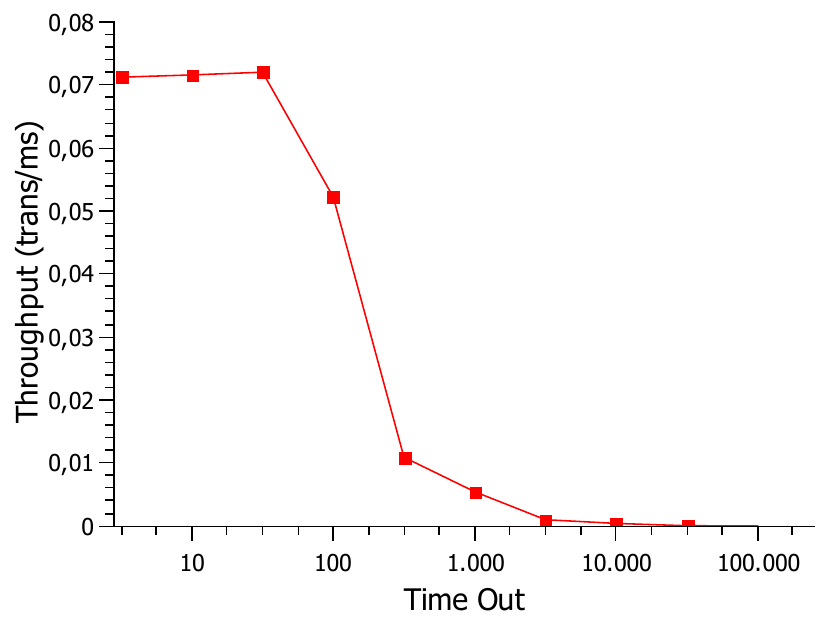}
  \caption{Flow - Changing Timeout}
  \label{fig:Tim_TP}
\end{subfigure}
\vspace{-0.2cm}
\caption{Case Study 02 Results Varying Block Size and Timeout}
\label{fig:sc2}
\end{figure*}

In our second case study, our objective is twofold. Initially, we adjust the block size while maintaining a substantial \textit{timeout} (\texttt{TIME\_OUT} = 10000ms). Subsequently, we modulate the \textit{timeout}, keeping the \textit{block size} constant at (\texttt{BLOCK} = 10). The intention behind this study is to methodically discern the individual ramifications of each parameter on the system's behavior. For this analysis, we standardized the arrival rate at 0.1 req/ms and adhered to the parameters delineated in Table~\ref{tab:cenarioC}. The outcomes of this investigative endeavor are illustrated in Figure \ref{fig:sc2}.

Analyzing Figures \ref{fig:sc2}a and \ref{fig:sc2}b reveals that system utilization increases with block size and \textit{timeout} adjustments, with \textit{timeout} changes exerting a stronger impact. This leads to a decrease in committer utilization due to bottlenecks, more so with \textit{timeout} alterations that drain resources faster than block size modifications. Figures \ref{fig:sc2}c and \ref{fig:sc2}d demonstrate varying trigger rates, identifying critical inflection points at a block size of 7 and a \textit{timeout} of 5000ms, beyond which discard rates and mean response times (MRT) significantly increase, with MRT oscillating between 1000ms and 25000ms and discard rates reaching up to 80\%. Throughput trends, as seen in Figures \ref{fig:sc2}i and \ref{fig:sc2}j, mirror the decline in committer utilization, highlighting system inefficiencies with an initial throughput of 0.07req/ms against an arrival rate of 0.1 req/ms. This analysis underscores the critical impact of block size and \textit{timeout} on system performance, indicating pronounced bottlenecks, especially in the ordering phase. A block size increment to 7 or a \textit{timeout} beyond 5000ms marks a significant turning point, where even minor changes can dramatically influence MRT, showcasing the delicate balance required in configuring these parameters.

\subsection{Case Study 03 - Dissecting the Interplay between Block Size and Timeout}

In our previous analysis, we separately evaluated the impact of block size and \textit{timeout} on system performance metrics, maintaining a \textit{timeout} of 10000ms to nullify its trigger rate while altering block size. Now, we shift focus to varying the \textit{timeout} across the same range, with the block size fixed at a lower value (\texttt{BLOCK} = 6), to uncover the intricate dynamics between these two parameters. Our objective is to understand how changes in \textit{timeout} influence both the \textit{timeout} and block activation rates, aiming to reveal their complex interplay as depicted in Figure \ref{fig:Toff}.

The graph shows a clear intersection of two distinct trajectories as the \textit{timeout} parameter increases, leading to a noticeable decrease in the activation rate per \textit{timeout}. This trend suggests that a longer \textit{timeout} period makes triggering by \textit{timeout} less frequent, increasing the chances of forming a complete block. The challenge of precisely adjusting these interrelated parameters, due to their complex interactions, has been underlined by researchers like Sukhwani et al. \citep{sukhwani2018performance}. Consequently, the model in this study provides crucial insights, enabling more accurate predictions of blockchain network behavior.
From our third case study, key findings include:

\begin{enumerate}
    \item The model effectively identifies the point where the formation of complete blocks overtakes partial blocks.
    \item A consistent trigger rate for blocks indicates the system has reached a high queuing state.
    \item Setting \texttt{TIME\_OUT} to 10ms leads to primarily partial blocks.
    \item A \texttt{TIME\_OUT} of 10000ms ensures the formation of complete blocks with 6 transactions each.
\end{enumerate}

This underscores the challenge of managing complete and partial block formations, affecting transaction time predictability. Blockchain network administrators must carefully balance block size and \textit{timeout} settings against transaction rates.

\subsection{Case Study 04 - Sensitivity Analysis}

\begin{figure}[!htb]
\centering
\begin{subfigure}[b]{0.5\linewidth}
\centering
\includegraphics[width=1.0\linewidth]{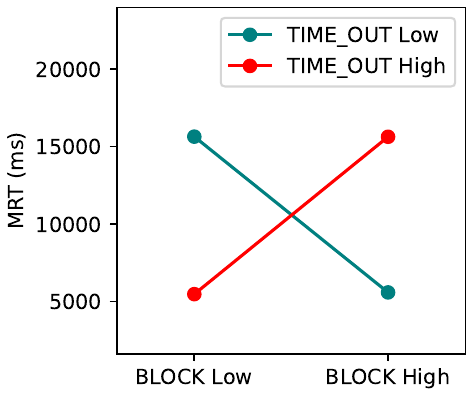}
\caption{Strength plot for Block x Timeout relationship.}
\label{fig:figura_com_subfiguras}
\end{subfigure}%
\hfill
\begin{subfigure}[b]{0.5\linewidth}
\centering
\includegraphics[width=1.0\linewidth]{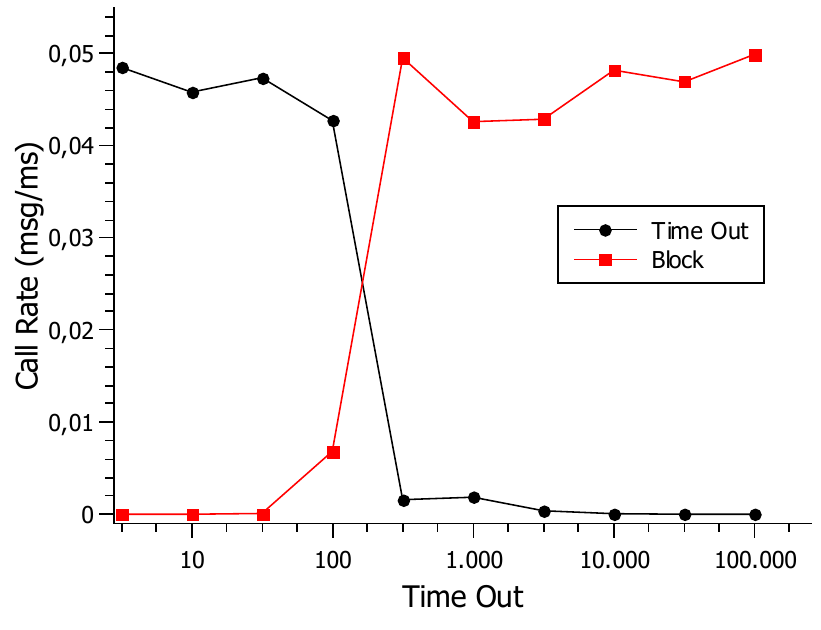}
\caption{Interaction between Block and Timeout}
\label{fig:Toff}
\end{subfigure}%
\hfill
\begin{subfigure}[b]{1.0\linewidth}
\centering
\includegraphics[width=0.65\linewidth]{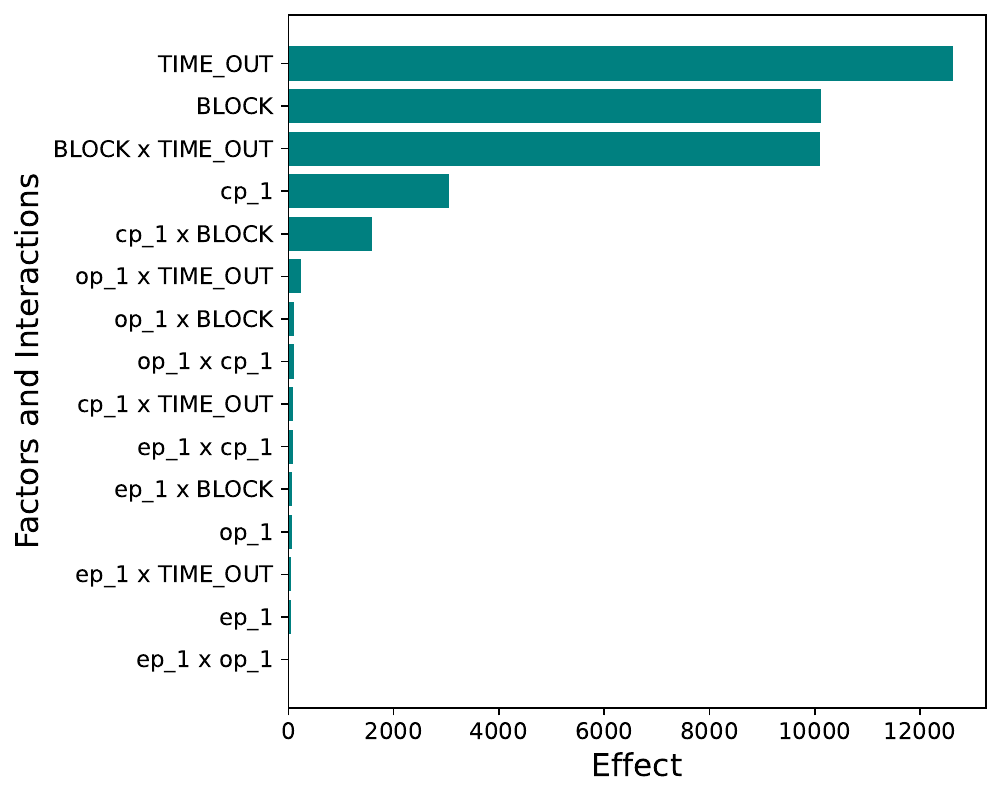}
\caption{Factors and Interactions and their respective effect.}
\label{fig:mod}
\end{subfigure}%
\caption{Comparative analysis of Block x Timeout relationship, factors, and interactions.}
\end{figure}

This case study uses a $2^k$ Design of Experiments (DoE) for Sensitivity Analysis to identify the key factors and their interactions affecting Mean Response Time (MRT), guided by previous insights. An effect plot in Figure \ref{fig:mod} shows the impact of different variables on MRT. Initial findings reveal \texttt{TIME\_OUT} as the most influential factor, with \texttt{block size} and its interaction with \texttt{TIME\_OUT} also significantly impacting system performance.

The values in this sensitivity analysis varied from \textbf{low} to \textbf{high}, indicating a minimum and a maximum value for each factor ($2^k$). Figure \ref{fig:figura_com_subfiguras} presents the strength plot for the relationship the standout, which is the relationship between block and TIME\_OUT.

The final case study reveals that the relationship between \textit{timeout} and block size significantly impacts system performance, with Mean Response Time (MRT) potentially escalating from 5000ms to over 15000ms, a 200\% increase. Such delays, far exceeding contemporary expectations for transaction speeds, could adversely affect user experience, with stakeholders possibly preferring public blockchains for their efficiency and lower operational costs.

However, the preference for private or permissioned blockchains among many governmental and industrial entities, driven by the need for strict data governance and access control, remains strong. This necessitates improvements in private blockchain technology to meet performance expectations. Our model serves as a crucial tool for addressing these challenges, helping to optimize blockchain configurations for enhanced performance and user satisfaction.

\section{Remarks}
~\label{sec:conclusion}

This paper introduces an SPN model designed to analyze the performance of permissioned blockchains, specifically within the Hyperledger Fabric framework. The model evaluates key performance indicators such as Mean Response Time (MRT), Throughput (TP), and Utilization, and assesses the impact of blockchain parameter changes—block size, \textit{timeout}, and transaction arrival rates—on the likelihood of transaction discards.

Through various parameter adjustments, including resource capacities and queuing dynamics, the model identifies critical areas for performance improvement. Four case studies demonstrate how blockchain configurations and the computational capacities of system components affect network latency and throughput, offering insights into optimizing Hyperledger Fabric's performance.

Looking ahead, plans include enhancing the model to incorporate weighted decision-making for protocol steps distribution among network machines based on probabilistic heuristics. This future direction aims to further dissect performance implications, potentially guiding subsequent case studies.

\bibliographystyle{plainnat}
\bibliography{refs}

\begin{thebibliography}{15}
\providecommand{\natexlab}[1]{#1}
\providecommand{\url}[1]{\texttt{#1}}
\expandafter\ifx\csname urlstyle\endcsname\relax
  \providecommand{\doi}[1]{doi: #1}\else
  \providecommand{\doi}{doi: \begingroup \urlstyle{rm}\Url}\fi

\bibitem[Aguiar et~al.(2022)Aguiar, dos Santos, Meneguette, Robson, and Ueyama]{aguiar_fgcs2022}
Erikson~J Aguiar, Alyson~J dos Santos, Rodolfo~I Meneguette, E~Robson, and J{\'o}~Ueyama.
\newblock A blockchain-based protocol for tracking user access to shared medical imaging.
\newblock \emph{Future Generation Computer Systems}, 134:\penalty0 348--360, 2022.

\bibitem[Androulaki and {et al.}(2018)]{androulaki_eurosys2018}
Elli Androulaki and {et al.}
\newblock {Hyperledger Fabric: A Distributed Operating System for Permissioned Blockchains}.
\newblock In \emph{Proc. of the EuroSys Conference}, 2018.

\bibitem[Greve et~al.(2018)Greve, Sampaio, Abijaude, Coutinho, Brito, and Queiroz]{greve_minicurso_sbrc2018}
Fabíola Greve, Leobino Sampaio, Jauberth Abijaude, Antonio~Augusto Coutinho, Italo Brito, and Silvio Queiroz.
\newblock {Blockchain e a Revolução do Consenso sob Demanda}.
\newblock In \emph{Proc. of SBRC Minicursos}, 2018.

\bibitem[Jain(1990)]{jainart}
Raj Jain.
\newblock \emph{The art of computer systems performance analysis: techniques for experimental design, measurement, simulation, and modeling}.
\newblock John Wiley \& Sons, 1990.

\bibitem[Jiang et~al.(2020)Jiang, Chang, Liu, Mi{\v{s}}i{\'c}, and Mi{\v{s}}i{\'c}]{jiang_springer_p2pna2020}
Lili Jiang, Xiaolin Chang, Yuhang Liu, Jelena Mi{\v{s}}i{\'c}, and Vojislav~B Mi{\v{s}}i{\'c}.
\newblock Performance analysis of hyperledger fabric platform: A hierarchical model approach.
\newblock \emph{Peer-to-Peer Networking and Applications}, 13:\penalty0 1014--1025, 2020.

\bibitem[Ke and Park(2022)]{ke_springer_cc2022}
Zuqiang Ke and Nohpill Park.
\newblock Performance modeling and analysis of hyperledger fabric.
\newblock \emph{Cluster Computing}, pages 1--19, 2022.

\bibitem[Maciel et~al.(2017)Maciel, Matos, Silva, Figueiredo, Oliveira, F{\'e}, Maciel, and Dantas]{maciel2017mercury}
Paulo Maciel, Rubens Matos, Bruno Silva, Jair Figueiredo, Danilo Oliveira, Iure F{\'e}, Ronierison Maciel, and Jamilson Dantas.
\newblock Mercury: Performance and dependability evaluation of systems with exponential, expolynomial, and general distributions.
\newblock In \emph{Proc. of PRDC}, pages 50--57. IEEE, 2017.

\bibitem[Melo et~al.(2021)Melo, Dantas, Pereira, and Maciel]{melo_supercomp2021}
Carlos Melo, Jamilson Dantas, Paulo Pereira, and Paulo Maciel.
\newblock Distributed application provisioning over ethereum-based private and permissioned blockchain: availability modeling, capacity, and costs planning.
\newblock \emph{The Journal of Supercomputing}, 77\penalty0 (9):\penalty0 9615--9641, 2021.

\bibitem[Melo et~al.(2022)Melo, Araujo, Dantas, Pereira, and Maciel]{melo_computing2022}
Carlos Melo, Jean Araujo, Jamilson Dantas, Paulo Pereira, and Paulo Maciel.
\newblock A model-based approach for planning blockchain service provisioning.
\newblock \emph{Computing}, 104\penalty0 (2):\penalty0 315--337, 2022.

\bibitem[R3(2020)]{spunta_2020}
R3.
\newblock Spunta case study. fast and transparent interbank reconciliation powered by distributed ledger technology.
\newblock \url{https://www.r3.com/wp-content/uploads/2020/11/Corda_Spunta_Case_Study_R3_Nov2020.pdf}, 2020.
\newblock Accessed: 2022-07-06.

\bibitem[Sukhwani et~al.(2018)Sukhwani, Wang, Trivedi, and Rindos]{sukhwani2018performance}
Harish Sukhwani, Nan Wang, Kishor~S Trivedi, and Andy Rindos.
\newblock Performance modeling of hyperledger fabric (permissioned blockchain network).
\newblock In \emph{Proc. of NCA}, pages 1--8. IEEE, 2018.

\bibitem[Wu et~al.(2022)Wu, Li, Liu, Zhang, Zhou, and Lu]{wu_acm_ease2022}
Ou~Wu, Shanshan Li, Liwen Liu, He~Zhang, Xin Zhou, and Qinghua Lu.
\newblock Performance modeling of hyperledger fabric 2.0.
\newblock In \emph{Proceedings of the International Conference on Evaluation and Assessment in Software Engineering 2022}, pages 357--365, 2022.

\bibitem[Xu et~al.(2021)Xu, Sun, Luo, Cao, Yu, and Vasilakos]{xu_ipm2021}
Xiaoqiong Xu, Gang Sun, Long Luo, Huilong Cao, Hongfang Yu, and Athanasios~V Vasilakos.
\newblock Latency performance modeling and analysis for hyperledger fabric blockchain network.
\newblock \emph{Information Processing \& Management}, 58\penalty0 (1):\penalty0 102436, 2021.

\bibitem[Xu et~al.(2019)Xu, Weber, and Staples]{xu_springer_book2019}
Xiwei Xu, Ingo Weber, and Mark Staples.
\newblock \emph{Architecture for blockchain applications}.
\newblock Springer, 2019.

\bibitem[Yuan et~al.(2020)Yuan, Zheng, Xiong, Zhang, and Lei]{yuan2020performance}
Pu~Yuan, Kan Zheng, Xiong Xiong, Kuan Zhang, and Lei Lei.
\newblock Performance modeling and analysis of a hyperledger-based system using gspn.
\newblock \emph{Computer Communications}, 153:\penalty0 117--124, 2020.

\end{thebibliography}

\end{document}